# Temporal and Spectral Shaping of Broadband THz Pulses in a Photoexcited Semiconductor

Mostafa Shalaby,[1,2*] Marco Peccianti,[3,] David G. Cooke,[4] Christoph P. Hauri,[2,5] and Roberto Morandotti[1*]

[1]*INRS-EMT, Varennes, Quebec J3X 1S2, Canada*

[2]*SwissFEL, Paul Scherrer Institute, 5232 Villigen PSI, Switzerland*

[3]*Department of Physics and Astronomy, University of Sussex, Pevensey Building II, 3A8, Falmer, Brighton BN1 9QH, United Kingdom*

[4]*Department of Physics, McGill University, 3600 rue University, Montréal, Québec H3A 2T8, Canada*

[5]*Ecole Polytechnique Federale de Lausanne, 1015 Lausanne, Switzerland*

 (*most.shalaby@gmail.com and morandotti@emt.inrs.ca)

Transmission through photoexcited semiconductors is used to temporally and spectrally shape a Terahertz pulse. By adjusting the optical pump-THz probe delay, we experimentally introduce a polar asymmetry in the pulse profile as high as 92%. To shape the spectrum, we apply the same technique after strongly chirping the Terahertz pulse. This led to significant reshaping of the spectrum with a 52% upshift of spectrum median. Pulse shaping techniques introduced here are of particular importance for temporal and spectral shape-sensitive THz nonlinear experiments.

Nonlinear Terahertz (THz) spectroscopy is gaining increasing interest thanks to its several potential applications in Physics, Chemistry, Biology, etc [1-6]. A number of THz-induced nonlinear mechanisms depend more on spectral content or the envelope function than the field profile of the pulse [2]. However, in some relevant cases the output field shape from THz sources needs to be tailored to suit the requirements of a particular experiment. For example, in ferroelectric materials, properly shaped THz pulses can coherently guide ions over a collective microscopic path [6]. A half cycle THz pulse (a pulse with a significant asymmetry between the amplitudes of the positive and negative parts of the oscillating fields) is shown to induce molecular reorientation [7]. Technologies to obtain intense THz generation have developed rapidly over the past few years with field strengths peaking on the MV/cm levels from both laser-based systems [9-12] and linear accelerators [13]. In addition, this is supported by several advances in field enhancement [14, 15] and polarization manipulation [16-17].

THz pulse shaping has been addressed in literature by manipulating the generation process [18-21]. However, shaping the pulse during generation has been mainly addressed in the generation with photoconductive antennas and periodically poled Lithium Niobate. Such sources are generally not suitable for nonlinear THz experiments due to the typical low peak fields. In a recent report, Sato et al, demonstrated strong control on the generated THz pulse by means of shaping the pulse generation. Yet this technique is strongly limited by the damage of the pulse shaper. They showed a field of 0.3 kV/cm [22]. Pulse shaping has also been achieved through linear filtering of a freely propagating THz in masks and waveguides [23,24].

In this work, we present a tunable THz pulse shaping technique operating in the time domain, capable of tailoring the temporal and spectral wave contents. Our technique operates via nonlinear excitation of free carriers in semiconductors by means of the optical pump–THz probe technique.

The transmitted THz field ($E_t$) through a photoexcited semiconducting layer of thickness $d$ much smaller than the THz wavelength is given by $E_t = (2Y_0 E_i - Jd)/(Y_0 + Y_S)$ where $E_i$ is the incident field. $Y_0$ and $Y_s$ are the admittance of free space and sample respectively. $J = n\,e\,v$ is the electron current density with $n$, $v$, and $e$ being the electron density, charge, and velocity in a respective order [25]. An increase in the charge density upon optical excitation is thus accompanied by an increase in the current density and consequently an attenuation of the transmitted THz pulse. A strong enough optical excitation can shield the THz. If, however, the probe-pump delay τ (when τ>0 the pump impinges before the probe transmission) is carefully selected, significant shaping of the THz time profile can be obtained as we show here. Temporal shaping of the broadband THz pulses considered here can be intuitively thought of as a process of clipping the pulse with the induced-carriers transition curve.

By simply modeling the charge-induced reflection, the Gabor-limit imposes a limit on the observable modulation time constant and the field spectral contents. We assume that no frequency products originate from the process. As a result, in a transform-limited pulse, no significant change in the temporal profile nor spectrum should be perceived for a given charge induced attenuation. To shape the spectrum, we deliberately induce significant chirp in the THz pulse. As different spectral components start to spread over different time points, carrier-induced modulation presented here is shown to change the spectrum distribution.

Measurements in this work were performed using a time resolved optical pump/THz probe scheme. Figure 1(a) shows a schematic diagram of the setup where the energy of a 35 fs pulse train (centered at wavelength λ=800 nm and with a repetition rate of 2.5 kHz) is split between the optical pump, the THz generation and the THz detection supply lines. Generation and detection were performed using optical rectification and electro-optical sampling, both using ZnTe crystals. The photoexcited layer is created on the surface of a 2 mm-thick high resistivity silicon wafer. Fig. 1(b) shows the waveform and the spectrum of the THz pulse transmitted through the silicon wafer. In Fig. 1(b) $A_1$, $A_2$, and $A_3$ designate the main wave peaks amplitudes. τ is the delay between the optical pump and the THz signal, assuming τ=0 corresponds to $A_2$ being reduced by a factor of $\sqrt{2}$. In Fig. 1(b) the sample is shined with an 695 μJ/cm$^2$ optical pulse arriving largely after the THz pulse. As we operate in a strong carrier excitation regime, a weak THz pulse attenuation is observed because the carrier lifetime is comparable with the period of the pump pulse train.

The THz pulse shown here is the typical one generated from optical rectification in a ZnTe crystal. A primary objective here is to increase the asymmetry in these fields i.e. $|A_1/A_2| \gg 1$ and $|A_1/A_3| \gg 1$. Such asymmetry is hard-to-achieve during the generation process but fundamental for many THz-nonlinear experiments. For example, when an oscillating symmetric pulse is to be used to trigger the magnetization switching in a magnetic material, the effective torque on the magnetic moment depends on the orientation of the applied magnetic field [5, 26]. As results a field-symmetric pulse produces a weak net switching effect, and often leads to a non-deterministic dynamics of the magnetization.

An asymmetric THz pulse is, therefore, highly desirable for such and other applications. Asymmetry is obtained here by varying τ such that $A_2$ and $A_3$ are gradually reduced in comparison.

Figure 1(c) shows the amplitude transition curve where the peak $A_2$ is shown against the delay τ over a period of 9 ps. $A_2$ is attenuated by > 90% when the THz arrives right after the optical pump. Pulse asymmetry is thus expected to change as τ is varied close to the temporal overlap between the probe and the pump. The asymmetry build-up process is illustrated in Fig. 1(d) where the original and modulated pulses are shown (in normalized units) for different τ points. As the delay is varied with the THz pulse getting overlapped by the optical one, (the most delayed) $A_3$ starts to get attenuated first, followed by $A_2$.

To evaluate the efficiency of the shaping process we calculate the temporal intensity modulation (shaping) depth:

$$M_{i1} = \frac{(A_{i1}^\circ)^2 - (A_{i1}^\circ)^2}{(A_{i1}^\circ)^2}, \quad i = 2, 3 \quad (1)$$

where $A_{21}$ and $A_{31}$ are the asymmetry factors given by $A_{21}=A_2/A_1$ and $A_{31}=A_3/A_1$, respectively. Circle superscripts denote measurement on the original unmodulated pulse. The attenuation, shown in Fig. 1(d) for few τ points, is mapped using equation (1) into $M_{21}$ and $M_{31}$ and is shown in Fig. 2(a). Modulation depths as high as 87% and 92%, respectively, are obtained. Such high modulation corresponds to a strong asymmetry being introduced in the THz pulse. In a typical application, the modulation process is first examined on the given pulse. Then, the desired asymmetry is obtained by selecting the right delay.

Finally, the aforementioned modulation process is centered about changing the pump-probe delay for a certain optical fluence. At a specific delay, changing the fluence leads to different electron current densities and thus the introduced THz attenuation and pulse asymmetry change accordingly. Figure 2(b) shows the transition curve of $A_2$ for different optical fluences that represents another degree of control over the pulse asymmetry.

As highlighted at the beginning of our letter, a transform-limited pulse should not undergo significant spectral change upon optical pumping. THz pulses generated using optical rectification in ZnTe are in general slightly chirped due to the chromatic dispersion of the phase mismatch with the optical pump in the rectification process. This, in turn, leads to small spectral changes during the temporal shaping process introduced so far. Indeed we can achieve significant spectral shaping by chirping the THz pulse before the time *slicing* process. We chirped the pulse by first letting it first propagate through a copper tube of length 278.3 mm. The inner diameter is 27.1 mm and results into an anomalous group velocity dispersion estimated at 1 THz to be $\beta_2$ = -1.8 ps$^2$/m where higher order dispersions have smaller contributions. Different frequency components are therefore spread over different delays and time slicing process can be used to filter parts of them and shift the center of the spectrum. As the pulse is negatively chirper, high frequencies tend to survive the clipping, whereas as $\tau$ decreases, the output spectrum is progressively enriched with lower frequency contributions. Figure 3(b) and 3(c) show the shaped time profiles and the corresponding spectra, respectively. Four sections are taken out to Fig. 3(d) and 3(e). As shown, the time clipping is accompanied by significant frequency upshift of the spectrum. The change in the spectrum is clearly appreciated in Fig. 3(f) where both the spectrum center ($f_2$) and the FWHM edges' frequencies ($f_1$ and $f_3$) are shown. All the spectra shown in Fig. 3 are in normalized units. This process is accompanied by [i] narrowing of the FWHM bandwidth and [ii] pushing the spectrum center up in frequency. Finally, the spectral modulation parameter $M_f = 1 - (f_1)^2 / (f_1^\circ)^2$ is shown in Fig. 3(g) where modulation as high as 52% was obtained. It should be noted that for a p-polarized THz impinging on the sample with a Brewster angle, the reflection exhibits the complementary effect, *i.e.* the spectrum median is downshifted. Alternative a low-pass filtering can be achieved by positively chirping the input pulse [27].

In conclusion, we have shown that optical pump-THz probe of carriers in semiconductors can be used to temporally and spectrally shape a THz pulse. Significant asymmetry in field polarity and shift in spectrum median were demonstrated. We believe that our results will help overcome some pulse shape limitations in nonlinear THz experiments.

This work was supported by the Canadian FQRNT and the NSERC. M.S. acknowledges an FQRNT MELS scholarship. M.S. wishes to thank Prof. Francois Legare (INRS-EMT) for helpful discussions on THz pulse shaping.

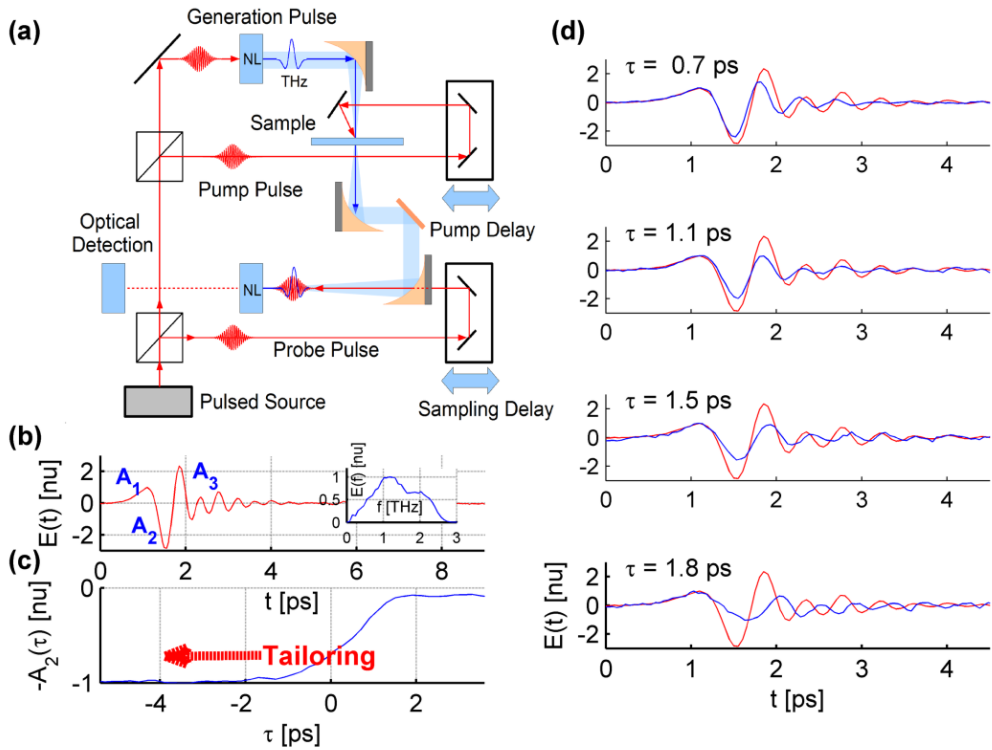

Fig.1 (a) schematic diagram of the optical pump / THz probe setup. (b) Time profile and spectrum of the transmitted THz pulse with the THz arriving before optical excitation. (d) $A_2$ amplitude transition as $\tau$ is varied. (d) THz amplitude transition shifted from the reference delay window shown in (c) and referred to here as $\tau_0$. (e) The corresponding transmitted THz pulse (blue) for different delay points shown along with the original unmodulated pulse (red).

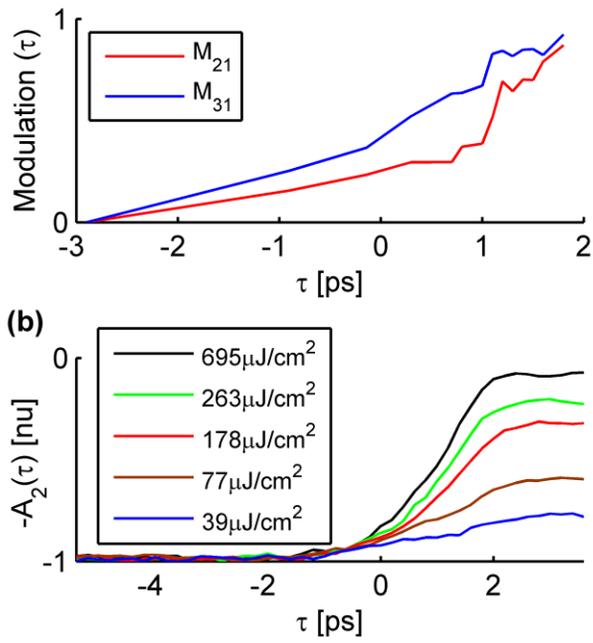

Fig.2 (a) intensity modulation depths for asymmetry factors $A_{21}$ and $A_{31}$. (b) $A_2$ amplitude transition for different fluence levels.

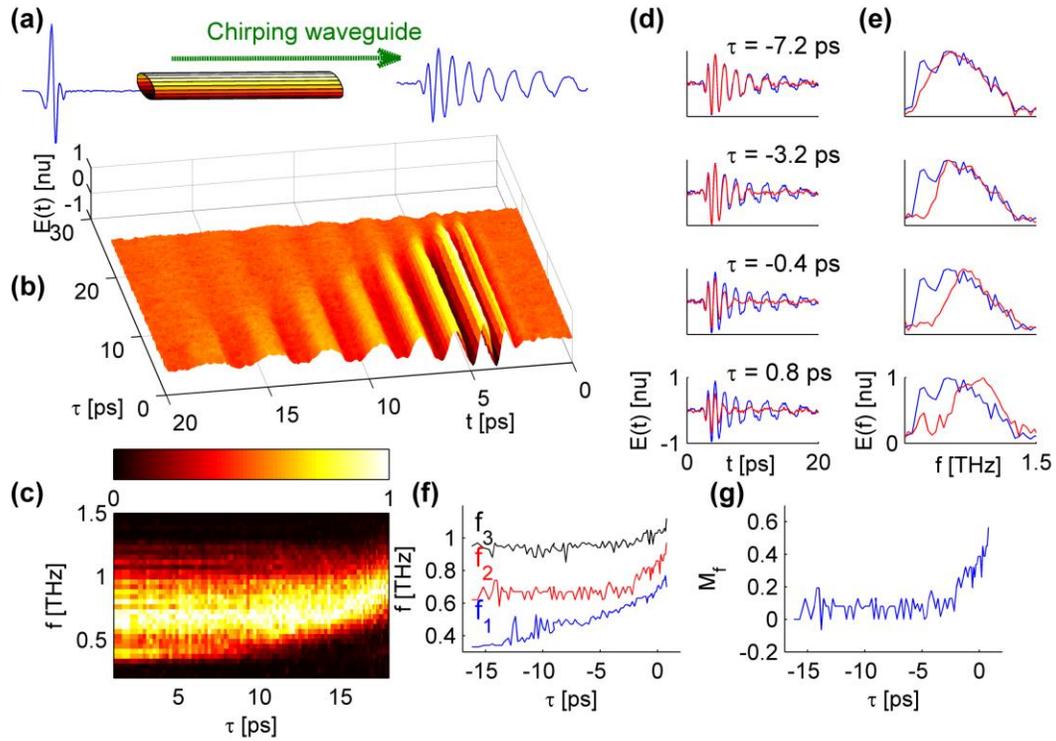

Fig.3 (a) Negative chirping of a THz pulse. (b) Time profile and (c) spectrum of the transmitted THz pulse as the pump-probe delay is varied. An arbitrary zero delay point is chosen. (d) Time profiles and (e) spectra of the modulated and unmodulated pulses at few delay points. (f) Trace lines of FWHM frequency edges and spectrum median. (g) Spectral modulation parameter.